\documentclass[twoside, twocolumn, journal, 10pt]{IEEEtran}
\IEEEoverridecommandlockouts
\usepackage{xcolor}
\usepackage{array}
\usepackage{tikz}
\usepackage{cite}
\usepackage[section]{placeins}
\usepackage{bm}
\usepackage{subcaption}
\usepackage{float}
\usepackage{amsmath}
\usepackage{graphicx}
\usepackage{epstopdf}
\usepackage{amssymb}
\usepackage{amssymb,amsmath}
\usepackage{amsgen,amsfonts,amsbsy,amsthm}
\usepackage{latexsym}
\usepackage{algorithm}
\usepackage{algpseudocode}
\usepackage[T1]{fontenc}
\allowdisplaybreaks
\makeatletter
\def\BState{\State\hskip-\ALG@thistlm}
\makeatother
\DeclareMathOperator*{\argmin}{arg\,min}
\DeclareMathOperator*{\argmax}{arg\,max}
\newcounter{defcounter}
\newcommand{\relu}[1]{\left(#1\right)^+}

\newcommand{\bvec}[1]{\boldsymbol{#1}}
\newcommand{\real}{\mathbb{R}}
\newcommand{\support}{\texttt{supp}}
\newcommand{\proj}[1]{\mathbf{P}_{#1}}
\newcommand{\dualproj}[1]{\mathbf{P}_{#1}^\perp}
\newcommand{\opnorm}[2]{\left\|#1\right\|_{#2}}
\newcommand{\norm}[1]{\left\|#1\right\|_2}
\newcommand{\abs}[1]{\left|#1\right|}

\newcommand{\expect}[1]{\mathbb{E}\left[#1\right]}
\newcommand{\prob}[1]{\mathbb{P}\left(#1\right)}

\newtheorem{lem}{Lemma}[section]
\newtheorem{thm}{Theorem}[section]

\newtheorem{prop}{Proposition}[section]
\newtheorem{define}{Definition}[section]
\title{Multiple Choice Hard Thresholding Pursuit (MCHTP) for Simultaneous Sparse Recovery and Sparsity Order Estimation}
\begin{document}
\vspace{-5mm}
\author{Samrat Mukhopadhyay and Himanshu Bhusan Mishra
\thanks{Samrat Mukhopadhyay$^1$ is supported by the FRS(167)/2021-2022/ECE Grant from IIT (ISM) Dhanbad,\\
     The authors are with the department of Electronics
	Engineering, Indian Institute of Technology (Indian School of Mines), Dhanbad, INDIA.
}}
%
\maketitle
\begin{abstract}
We address the problem of sparse recovery using greedy compressed sensing recovery algorithms, without explicit knowledge of the sparsity. Estimating the sparsity order is a crucial problem in many practical scenarios, e.g., wireless communications, where exact value of the sparsity order of the unknown channel may be unavailable a priori. In this paper we have proposed a new greedy algorithm, referred to as the Multiple Choice Hard Thresholding Pursuit (MCHTP), which modifies the popular hard thresholding pursuit (HTP) suitably to iteratively recover the unknown sparse vector along with the sparsity order of the unknown vector. 
We provide provable performance guarantees which ensures that MCHTP can estimate the sparsity order exactly, along with recovering the unknown sparse vector exactly with noiseless measurements. The simulation results corroborate the theoretical findings, demonstrating that even without exact sparsity knowledge, with only the knowledge of a loose upper bound of the sparsity, MCHTP exhibits outstanding recovery performance, which is almost identical to that of the conventional HTP with exact sparsity knowledge. Furthermore, simulation results demonstrate much lower computational complexity of MCHTP compared to other state-of-the-art techniques like MSP. 
\end{abstract}
\section{Introduction}
\label{sec:intro}
Sparse recovery is a signal processing technique of significant research interest in diverse practical problems, ranging from wireless communication, image processing, radar signal processing, to name a few. At the core of any sparse recovery problem lies the ill-posed inverse problem of solving an under-determined system of linear equations. However, the crux of addressing such ill-posed problems lies in assuming that the unknown vector is \emph{sparse}, i.e., many of its entries are $0$. Mathematically, the problem becomes the following: 
\begin{align}
\label{eq:cs-problem}
    \begin{aligned}
    \min_{\bvec{x}\in \real^N:\|\bvec{x}\|_0\le K}\norm{\bvec{y}-\bvec{\Phi x}}^2,
    \end{aligned}
\end{align}
where $\opnorm{\bvec{x}}{0}$ is the \emph{sparsity} of $\bvec{x}$, denoting the number of non-zero entries of ${\bvec{x}}$, and $K$ is an upper bound of the sparsity. The formulation~\eqref{eq:cs-problem} is at the cornerstone of the so called field of Compressed Sensing (CS)~\cite{donoho2006compressed} and a large body of research has been devoted obtaining fast and low complexity algorithms, enjoying optimal or sub-optimal recovery guarantees for the solution. The problem~\eqref{eq:cs-problem} is inherently difficult to solve as the constraint set, $\{\bvec{x}:\opnorm{\bvec{x}}{0}\le K\}$ is non-convex. Although convex relaxation approaches can be employed to obtain provably sub-optimal algorithms, they are often slow for large dimensions. Greedy algorithms provide an alternate route to solve such difficult problems. Although heuristic in nature, such methods provide fast alternative to convex relaxation, often with competitive and provably sub-optimal recovery guarantees. Some canonical examples of such greedy methods are: orthogonal matching pursuit (OMP)~\cite{pati1993orthogonal}, iterated hard thresholding (IHT)~\cite{blumensath2009iterative}, subspace pursuit (SP)~\cite{dai2009subspace}, compressive sampling matching pursuit (CoSAMP)~\cite{needell2009cosamp}, hard thresholding pursuit (HTP)~\cite{foucart2011hard}, to name a few. However, execution of many of these greedy methods require explicit knowledge of the exact sparsity of the unknown vector. While orthogonal matching pursuit does not strictly require to know the sparsity for its execution, an exact knowledge of sparsity provides better recovery guarantee. On the other hand, although the algorithms HTP, SP or CoSAMP can work with an upper bound of the sparsity, the computational complexity of these algorithms increase substantially if that upper bound is quite large. Furthermore, in many practical problems, e.g., Delay-Doppler (DD) path estimation in orthogonal time frequency space (OTFS) modulation, the knowledge of exact number of DD paths is difficult to come by, rendering most of the above greedy methods of little use in such contexts. This necessitates sparsity order estimation (SOE) as a crucial preprocessing for greedy sparse recovery algorithms.

The SOE problem has been addressed by several recent research works~\cite{wang2012sparsity,sharma2014compressive,gao2018sparsity,semper2018sparsity,ravazzi2018sparsity,thiruppathirajan2022sparsity}. However, most of these works have used either various heuristic methods~\cite{wang2012recovery,sharma2014compressive,gao2018sparsity} with no provable performance analysis, or statistical as well as methods based on asymptotic random matrix theory~\cite{semper2018sparsity,ravazzi2018sparsity,thiruppathirajan2022sparsity}, which lack provable non-asymptotic performance analysis for SOE. Furthermore, most of these works do not use greedy sparse recovery algorithms. One of the first works which has used a greedy sparse recovery algorithm in conjunction with unknown sparsity estimation is the modified subspace pursuit (MSP)~\cite{rasheed2020sparse}. MSP gradually increments the sparsity and runs the SP algorithm until convergence and stops only when the convergence error is small enough. However, since MSP requires SP to run repeatedly for many sparsity estimates, then it can suffer from significant computational burden. Graded HTP (GHTP)~\cite{bouchot2016hard} is another method which recovers a sparse vector without explicit knowledge of sparsity. However, GHTP does not provide an explicit estimate of the sparsity and rather the sparsity has to be looked up from the iteration number of the algorithm when the algorithm stops. Furthermore, the stopping criterion of GHTP is set in such a way that it can overestimate the sparsity significantly for large sparsity values. 

In this paper, we propose the multiple choice hard thresholding pursuit, abbreviated as MCHTP, which simultaneously estimates sparsity order as well as the unknown sparse vector until convergence. The salient features of our work are listed as below:
\begin{enumerate}
    \item To our knowledge, MCHTP is the first greedy method that can simultaneously provide both sparsity estimate as well as estimate of sparse vector in each iteration. This simultaneous execution helps in saving computational burden significantly.
    \item We theoretically provide a complete characterization of the evolution of the sparsity sequence estimated by MCHTP and provide theoretical bounds that ensure exact recovery for noiseless measurements.
    \item Our simulation results clearly demonstrate the efficacy of the proposed MCHTP in terms of fast sparsity order estimation performance and sparse recovery performance which is at par with the sparse recovery performance of HTP with sparsity knowledge.
\end{enumerate} 
%
%
%
\section{Notation}
The following notations have been used throughout the paper :`$\top$' in
superscript indicates transposition of matrices / vectors. The entries of a vector $\bvec{x}\in \real^N$ are denoted by $x_1,\cdots, x_N$.
%
For any $S\subseteq \{1,2,\cdots, N\}$,
$\bvec{x}_S$ denotes the vector $\bvec{x}$ restricted to $S$,
i.e., $\bvec{x}_S$ consists of those entries of $\bvec{x}$ that have indices belonging to $S$.
The operator $H_K(\cdot)$ returns the \emph{$K$-best approximation} of a vector, i.e., for any vector $\bvec{x}\in \real^N$, $H_K(\bvec{x})=\arg\min_{\bvec{z}\in \real^N:\opnorm{\bvec{z}}{0}\le K}\opnorm{\bvec{z} - \bvec{x}}{2}$. Similarly, $L_K(\bvec{x})$ returns the support of the $K$-best approximation of vector $\bvec{x}$. 
The symmetric difference $\Delta$, between two sets $A, B$, is defined as $A\Delta B: = (A\setminus B)\cup (B\setminus A)$.
\section{Proposed Algorithm}
\label{sec:proposed-algo}
\begin{algorithm}[t!]
\caption{Multiple Choice HTP (MCHTP)}
\label{algo:MCHTP}
\begin{algorithmic}[1]
\Require A sparsity upper bound $\bar{K}$, an initial estimate $\bvec{x}^0=\bvec{0}$ (so that $K_0=0$), measurement vector $\bvec{y}$, measurement matrix $\bvec{\Phi}$, step size $\mu>0$, number of iterations to run $T$ and a factor $\epsilon>0$.
\For {$t=1,\cdots,\ T$ }
\State Set $K_{t,0} = K_{t-1}$ and sample  $K_{t,1}$ uniformly randomly from the set $\{1,\cdots, \bar{K}\}\setminus \{K_{t,0}\}$.
\For {$i=0,1$}
\State $\Lambda_{t,i} = L_{K_{t,i}}\left(\bvec{x}^{t-1} + \mu\bvec{\Phi}^\top \left(\bvec{y} - \bvec{\Phi x}^{t-1}\right) \right)$
\State $\bvec{x}^t_{i} = \arg\min_{\bvec{z}:\support(\bvec{z})=\Lambda_{t,i}}\norm{\bvec{y}-\bvec{\Phi z}}$
\State $E_{t,i}=\norm{\bvec{y}-\bvec{\Phi x}^t_{i}}^2$
\EndFor 
\State $\Delta E_t = \abs{E_{t,1}-E_{t,0}}$
\If{$\Delta E_t> \epsilon$}
\State $i_t = \argmin_{i\in \{0,1\}}E_{t,i}$
\Else 
\State $i_t = \argmax_{i\in \{0,1\}}E_{t,i}$ 
\EndIf 
\State $K_t = K_{t,i_t}$, $\bvec{x}^t = \bvec{x}^t_{i_t}$
\EndFor
\end{algorithmic}
\end{algorithm}
%


The Multiple Choice HTP (MCHTP) algorithm is elaborated in Table~\ref{algo:MCHTP}. The main idea of MCHTP is to evaluate the HTP computations iteratively on two different choices of sparsity and select one using a suitable criterion. We explain this in the following paragraph.

At each iteration $t\ge 1$ of MCHTP, at step $2$, we begin with two guesses for the sparsity order, $K_{t,0}$ and $K_{t,1}$. We set $K_{t,0} = K_{t-1}$, where $K_{t-1}$ is the sparsity order estimated at the end of the last iteration $t-1$, whereas, $K_{t,1}$ is chosen uniformly randomly from the set $[\bar{K}]\setminus \{K_{t,0}\}$, where $[\bar{K}]:=\{1,2,\cdots, \bar{K}\}$. In steps $4-5$, one iteration of HTP is executed for each of these sparsity orders starting with the same initial estimated vector $\bvec{x}^{t-1}$ and two estimates $\bvec{x}^{t}_{i}, i=0,1$ are produced and the corresponding estimation errors $E_{t,i}, i=0,1$ are calculated in step $6$. At this step, it should be recalled that HTP ensures that the larger sparsity will yield smaller estimation error. These errors are now used to select sparsity order in the following way: always select the smaller sparsity (with larger error) unless the absolute difference in errors, i.e, $\Delta E_t = \abs{E_{t,1}-E_{t,0}}$ is larger than the predefined threshold $\epsilon$. In the former case, the estimation errors do not differ ``much'', i.e., by less than $\epsilon$, and the smaller sparsity corresponding to the slightly larger estimation error is chosen. In the latter case, the error corresponding to the smaller sparsity is to large to choose the corresponding sparsity and therefore, the larger sparsity with smaller error is estimated. 
\section{Convergence Analysis of MCHTP}
\label{sec:analysis-MCHTP}
In this section, we present a theoretical convergence analysis of the proposed MCHTP algorithm. For the sake of simplicity, we consider the noiseless measurement model $\bvec{y}=\bvec{\Phi x}.$ Furthermore, we consider $\mu = 1$, although the analysis can be generalized in a straightforward way for general $\mu>0$. Before embarking on the mathematical analysis, let us provide an intuitive outline of the analysis. We first specify a key result which states that if the sensing matrix satisfies certain conditions and the predefined threshold $\epsilon$ is chosen properly then an iteration $t$ of MCHTP satisfies $\Delta E_t>\epsilon$ as long as $\min_{i} K_{t,i}<K$, ensuring that $K_t = \max_i K_{t,i}$. This initiates the first phase of MCHTP consisting of a no-decreasing sequence of sparsity estimates $\{K_t\}$, as long as $K_t< K$. Once $K_t\ge K$, the phase two begins. In phase two, one always has $\max_iK_{t,i}\ge K$, although $\Delta E_t$ might fluctuate around $\epsilon$. In the third and final phase, $\Delta E_t$ is always smaller than $\epsilon$, ensuring that $\{K_t\}$ is a decreasing sequence, albeit not smaller than $K$. This ensures that the sparsity tracked by the algorithm gradually converges to $K$ from above. We also establish a decay inequality of the sequence of estimated vectors $\{\bvec{x}^{t-1}\}$, which provides, along with the estimated sparsity sequence $\{K_t\}$, a provable convergence guarantees for MCHTP.
We now state and prove below a crucial lemma about the evolution of the intermediate estimates $\bm{x}_i^t$: 
\begin{prop}
\label{prop:x^t-x^t-i-decay-inequalities}
At any iteration $t$, for each $i=0,1$, the estimate $\bvec{x}_i^t$ satisfies the following decay inequality:
\begin{align}
\label{eq:x^t_i-decay-inequality}
\norm{\bvec{x}^t_i - \bvec{x}} & \le \rho_{t,i}\norm{\bvec{x}^{t-1}-\bvec{x}} + \gamma_{t,i}\norm{\bvec{x}_{\overline{\Gamma_{t,i}}}},
\end{align}
while $\bvec{x}^t$ satisfies the following:
\begin{align}
\label{eq:x^t-decay-inequality}
\norm{\bvec{x}^t - \bvec{x}} & \le \rho_{t}\norm{\bvec{x}^{t-1}-\bvec{x}} + \gamma_{t}\norm{\bvec{x}_{\overline{\Gamma_{t}}}},
\end{align}
where,
\begin{eqnarray}
\label{eq:rho-t-i_rho-t-def}
\rho_{t,i} = \frac{\sqrt{2}\delta_{K_{t,i}+K_{t-1}+K}}{\sqrt{1-\delta_{K_{t,i}+K}^2}}, && \rho_{t} = \frac{\sqrt{2}\delta_{K_{t}+K_{t-1}+K}}{\sqrt{1-\delta_{K_{t}+K}^2}},\\
\label{eq:gamma-t-i_gamma-t-def}
\gamma_{t,i} = \frac{\sqrt{2}}{\sqrt{1-\delta_{K_{t,i}+K}^2}}, && \gamma_{t} = \frac{\sqrt{2}}{\sqrt{1-\delta_{K_{t}+K}^2}},\\
\end{eqnarray}
Moreover, $\Gamma_{t,i}$ (resp. $\Gamma_t$) is the support of the (magnitude-wise) top $K_{t,i}$ (resp. $K_t$) entries of $\bvec{x}$.
\end{prop}
\begin{proof}
An upper bound of $\norm{\bvec{x}^t_i-\bvec{x}}$ can be obtained by employing the analysis technique of HTP~\cite{foucart2011hard}. However, this is not straightforward as the sparsity estimates are different at different time instants and are in general unequal to the original sparsity. Therefore, we will modify the analysis of \cite{foucart2011hard} suitably to obtain the desired bounds. The detailed proof can be found in Appendix~\ref{sec:appendix-proof-prop-x^t-x^t-i-decay-inequalities}.
\end{proof}

We now state a central result which ensures that, for a suitable choice of $\epsilon$, if the sensing matrix satisfies certain condition, then the estimate sparsity sequence $\{K_t\}$ is non-decreasing in the Phase I as long as $\min_{i}K_{t,i}<K$.
\begin{prop}
\label{prop:phase-I-MCHTP}
Let us denote $\delta: = \delta_{2\bar{K}+K}$ which satisfies the following:
\begin{align}
\label{eq:delta-condition-lower-bound}
\delta & < \frac{1}{1+(3+5R)\sqrt{2K}}
\end{align}
where $R = \frac{x_{\max}}{x_{\min}}$, where $x_{\min}=\min\{\abs{x_j}:j\in \Lambda\},\ x_{\max}=\max\{\abs{x_j}:j\in \Lambda\}$.

At an iteration $t$, let $\min_{i\in\{0,1\}}K_{t,i}<K$. Then, $K_{t} = \max_{i\in \{0,1\}}K_{t,i}$ if the following is satisfied along with~\eqref{eq:delta-condition-lower-bound}: 
\begin{align}
\label{eq:epsilon-condition-phaseI}
0<\epsilon & <  \frac{(1-\delta)}{(1+\delta)^2}\left(a_K(\delta)x_{\min} - b(\delta)x_{\max}\right)^2,
\end{align}
where 
\begin{eqnarray}
\label{eq:a_K-b_K-def}
a_K(\delta) = \frac{1-\delta}{\sqrt{K}}-3\sqrt{2}\delta, &  b(\delta) = 5\sqrt{2}\delta.
\end{eqnarray}  
\end{prop}  
%
\begin{proof}
The idea of the proof is to show that if at any iteration $t$, $\min_{i\in \{0,1\}}\{K_{t,i}\}<K$, and if $\epsilon$ is \textit{small enough}, i.e., if $\epsilon$ satisfies the bound~\eqref{eq:epsilon-condition-phaseI}, then $\Delta E_t>\epsilon$. Then, by the step 11 of MCHTP (Algorithm~\ref{algo:MCHTP}), we have $K_t=\max_{i\in \{0,1\}}K_{t,i}$. The detailed proof can be found in Appendix~\ref{sec:appendix-proof-prop-phase-I-MCHTP}. 
\end{proof}
\noindent\textbf{Discussion}:
\label{rem:epsilon-phaseI}
The explicit dependence of the conditions~\eqref{eq:delta-condition-lower-bound} and~\eqref{eq:epsilon-condition-phaseI} on the signal structure through the relative signal magnitude, expressed by $R=\frac{x_{\max}}{x_{\min}}$ has interesting implications for different types of signals, e.g., flat, decaying, etc. We discuss below the effect of some the important signal structures on the conditions~\eqref{eq:delta-condition-lower-bound} and~\eqref{eq:epsilon-condition-phaseI}.  
\begin{enumerate}
    \item \textit{Flat signal structure}: In this case, the nonzero entries of $\bvec{x}$ are all of the same magnitude and differ only by sign, implying $x_{\min}=x_{\max} = \frac{\norm{\bvec{x}}}{\sqrt{K}}$ and $R=1$. Therefore, the condition for successful completion of phase I for flat signals in the noiseless setting is ensured by the following two conditions:
    \begin{align}
    \label{eq:delta-condition-flat}
        \delta & < \frac{1}{1+8\sqrt{2K}},\\
    \label{eq:epsilon-condition-phaseI-noiseless-flat}
        0<\epsilon & <  \frac{(1-\delta)\norm{\bvec{x}}^2}{K(1+\delta)^2}\left(a_K(\delta) - b(\delta)\right)^2.
    \end{align}
    \item \textit{Linear signal structure:} In this case, the nonzero entries follow a linear profile, i.e., the $j^\mathrm{th}$ largest entry is of the form $\alpha j$, for $j=1,\cdots, K$, where $\alpha = \sqrt{\frac{6\norm{\bvec{x}}^2}{K(K+1)(2K+1)}}$. Therefore, $x_{\min}=\alpha$ and $x_{\max} = K\alpha$, so that $R=K$. Consequently, the Phase I of MCHTP successfully completes if 
    \begin{align}
        \label{delta-condition-linear}
        \delta & < \frac{1}{1 + (3+5K)\sqrt{2K}},\\
        \label{eq:epsilon-condition-phaseI-noiseless-linear}
        0<\epsilon & < \frac{6(1-\delta)\norm{\bvec{x}}^2}{(1+\delta)^2K(K+1)(2K+1)}(a_K(\delta) - Kb(\delta))^2.
    \end{align}
    \item \textit{Decaying signal structure:} In this case, the nonzero entries follow a geometric profile so that, $x_{\min}=\alpha^{K-1}x_{\max}$, for some $\alpha\in (0,1]$. Therefore, $x_{\max} = \norm{\bvec{x}}\sqrt{\frac{1-\alpha^2}{1-\alpha^{2K}}}$, and $R=\alpha^{1-K}$. Consequently, the phase I of MCHTP is ensured to succeed in this case for noiseless setting if the following conditions are satisfied:
        \begin{align}
    \label{eq:delta-condition-decaying}
        \delta & < \frac{1}{1+(3+5\alpha^{1-K})\sqrt{2K}},\\
    \label{eq:epsilon-condition-phaseI-noiseless-decaying}
        0<\epsilon & <  \frac{(1-\delta)(1-\alpha^2)\left(a_K(\delta)\alpha^{K-1} - b(\delta)\right)^2}{K(1+\delta)^2(1-\alpha^{2K})}\norm{\bvec{x}}^2.
    \end{align}
    It can be easily observed that the conditions~\eqref{eq:delta-condition-decaying} and~\eqref{eq:epsilon-condition-phaseI-noiseless-decaying} are considerably more prohibitive than the conditions~\eqref{eq:delta-condition-flat} and~\eqref{eq:epsilon-condition-phaseI-noiseless-flat}. This indicates that successful completion of phase I of MCHTP might be much easier for flat signals than for the decaying signals.
\end{enumerate}

%
Proposition~\ref{prop:phase-I-MCHTP} is a significant tool to analyze the evolution of the sequence $\{K_t\}$. Using this, we proceed to obtain a characterization of the \emph{Phase I} of MCHTP. We first define the duration of the Phase I as below:
\begin{define}
\label{def:phase1-def}
The phase I of MCHTP is defined to consist of the time slots $\{0,1,\cdots, T_1-1\}$, where $T_1$ is a random time defined as below:
\begin{align}
    \label{eq:T1-def}
    T_1 & =\min\{t\ge 1:K_t\ge K\}.
\end{align}
\end{define}
We now proceed to provide a characterization of $T_1$.
\begin{lem}
\label{lem:T1-characterization}
Under the satisfaction of the conditions~\eqref{eq:delta-condition-lower-bound} and~\eqref{eq:epsilon-condition-phaseI}, the sequence $\{K_t\}$ is non-decreasing during Phase I and $K_t\ge K,\ \forall t\ge T_1$. Furthermore, $T_1$ is characterized as below:
\begin{align}
\label{eq:T_1-characterization}
\prob{T_1 = t} = \left\{
\begin{array}{ll}
	1 - p, & t = 1\\
	pq^{t-2}(1-q), & t> 1,
\end{array}
\right.
\end{align}
where $p=\frac{K-1}{\bar{K}}$ and $q = \frac{K-2}{\bar{K}-1}$.
\end{lem} 
\begin{proof}
To prove the first part of the claim, note that, by definition, during the phase I of MCHTP, we have $K_t<K$, so that $\min_{i\in \{0,1\}}K_{t,i}<K$. Therefore, by Proposition~\ref{prop:phase-I-MCHTP}, if the conditions~\eqref{eq:delta-condition-lower-bound} and~\eqref{eq:epsilon-condition-phaseI} are satisfied, then during phase I, $K_t = \max_{i\in \{0,1\}}K_{t,i},\ 1\le t\le T_1$. By the description of MCHTP in table~\ref{algo:MCHTP}, it follows that during Phase I of MCHTP, i.e., for $1\le t\le T_1$, we have 
\begin{align}
    K_t & = \max\{K_{t-1},K_{t,1}\} = \max\{\max\{K_{t-2},K_{t-1,1}\}, K_{t,1}\} \nonumber\\
    \ & = \max\{K_{t-2},K_{t-1,1},K_{t,1}\}\nonumber\\
    \ & = \cdots = \max\{K_0,K_{1,1},\cdots, K_{t,1}\} = \max_{1\le s\le t}K_{s,1},
\end{align}
since $K_0=0$ and $K_t\ge 1,\ \forall t\ge 1$. Therefore, $\{K_t\}$ is non-decreasing during Phase I of MCHTP.

To see that $K_{t}\ge K,\ \forall t\ge T_1$, we proceed via induction. First note that $K_{T_1}\ge K$ by definition. Then assume that it holds for $T_1,\cdots, t-1$ for some $t\ge T_1+1$. We have to prove that the claim holds for $t$. If $\min_{i\in \{0,1\}}K_{t,i}\ge K$, then trivially, $K_t \ge K$. So let us assume that $\min_{i\in \{0,1\}}K_{t,i}< K$. If the conditions~\eqref{eq:delta-condition-lower-bound} and~\eqref{eq:epsilon-condition-phaseI} are satisfied, it follows from Proposition~\ref{prop:phase-I-MCHTP} that $K_t=\max\{K_{t-1},K_{t,1}\}$. Note that by assumption $K_s\ge K$, for $s=T_1,\cdots, t-1$, so that $K_{t-1}\ge K$. Since we have assumed that $\min_{i\in \{0,1\}}K_{t,i} = \min{K_{t-1},K_{t,1}}<K$, we must have $K_{t,1}<K$ since $K_{t-1}\ge K$ by assumption. Therefore, $K_t=\max\{K_{t-1},K_{t,1}\}\ge K$. This completes the induction argument.

To obtain a characterization of $T_1$, note that as the sequence $\{K_t\}$ is non-decreasing in Phase I, it follows that $K_t=\max_{s=1,\cdots, t}K_{s,1}$. Therefore, it follows from the definition of $T_1$ that $T_1=\min\{t\ge 1: K_{t,1}\ge K\}$.
Therefore \begin{align}
	\prob{T_1 = 1} & = \prob{K_{1,1}\ge K} = \frac{\bar{K}-K+1}{\bar{K}}.
\end{align}
On the other hand, for $t> 1$,
\begin{align}
	\prob{T_1 = t} & = \prob{\max_{1\le s\le t-1}K_{s,1}< K, K_{t,1}\ge K}\nonumber\\
	\ & = \left(\frac{K-1}{\bar{K}}\right)\left(\frac{K-2}{\bar{K}-1}\right)^{t-2}\left(\frac{\bar{K}-K+1}{\bar{K}-1}\right).
\end{align}
The last expression follows since given $K_{t-1},\ t>1$, $K_{t,1}$ is sampled uniformly randomly from the set $\{1,\cdots, \bar{K}\}\setminus\{K_{t-1}\}$.
\end{proof}
Using the above result, we will now state and prove our first result on the convergence of MCHTP. 
\begin{thm}
\label{thm:convergence-MCHTP}
At any iteration $t$, the iterate $\bvec{x}^t$ satisfies the following decay inequality:
\begin{align}
\label{eq:x^t-decay-T1}
\norm{\bvec{x}^t - \bvec{x}} & \le \rho\norm{\bvec{x}^{t-1}-\bvec{x}} + \gamma\norm{\bvec{x}_{\overline{\Gamma_{t}}}},\ 1\le t<T_1,\\
\label{eq:x^t-decay-after-T1}
\norm{\bvec{x}^t - \bvec{x}} & \le \rho\norm{\bvec{x}^{t-1}-\bvec{x}},\ t\ge T_1,
\end{align}
where $\rho = \frac{\sqrt{2}\delta}{\sqrt{1-\delta^2}},\ \gamma = \frac{\sqrt{2}}{\sqrt{1-\delta^2}}$.
Consequently, if $\rho<1$, $\bvec{x}^t\to \bvec{x}$ if $T_1$ is finite with probability $1$.
\end{thm}
\begin{proof}
The proof relies on Proposition~\ref{prop:phase-I-MCHTP} as well as the Lemma~\ref{lem:T1-characterization}. The inequality~\eqref{eq:x^t-decay-inequality} trivially implies the inequality~\eqref{eq:x^t-decay-T1}. Now, by the Lemma~\ref{lem:T1-characterization}, $K_t\ge K$ for $t\ge T_1$. Therefore, by the definition of $\Gamma_t$ in Proposition~\ref{prop:phase-I-MCHTP} we have that $\Lambda\subset \Gamma_t,\ \forall t\ge T_1$, where $\Lambda$ is the true support of $\bvec{x}$. Consequently, for all $t\ge T_1$, $\norm{\bvec{x}_{\overline{\Gamma}_t}}=0$, which proves the inequality~\eqref{eq:x^t-decay-after-T1}. Finally, if $\rho<1$ and $T_1$ is finite with probability $1$ (which is true due to Lemma~\ref{lem:T1-characterization}), it is trivial to see that $\lim_{t\to\infty}\norm{\bvec{x}^t-\bvec{x}}=0\implies \bvec{x}^t\to \bvec{x}$.  
\end{proof}
On the completion of the first $T_1$ iterations, MCHTP enters its \textit{phase II} during which the estimated sparsity sequence might be increasing or decreasing, albeit remaining $\ge K$.
The following lemma is key to provide an estimate of the duration of the phase II of MCHTP. Before proceeding, we first define the Phase II formally.
\begin{define}
\label{def:phase-II}
The Phase II of MCHTP is defined to be the time duration consisting of the time slots $T_1,\cdots, T_1+T_2-1$, where
\begin{align}
    \label{eq:T_2-def}
    T_2 = \min\left\{t\ge 1: \{K_{s+T_1}\}_{s\ge t}\ \mbox{\emph{is a non-increasing sequence}}\right\}.
\end{align}
\end{define}
\begin{prop}
\label{prop:phase-2-duration}
If $\epsilon$ satisfies conditions~\eqref{eq:delta-condition-lower-bound} and \eqref{eq:epsilon-condition-phaseI} 
then the phase II of MCHTP consists of $T_2$ iterations, where 
\begin{align}
\label{eq:T_2-bound}
T_2 & \le \left\lfloor\relu{\frac{\ln\left[\frac{\sqrt{(1-\delta)\epsilon}}{(1+\delta)\norm{\bvec{x}}}\right]}{\ln \rho}}\right\rfloor,
\end{align}
where $\rho=\frac{\sqrt{2}\delta}{\sqrt{1-\delta^2}}$ and $\relu{x}=\max\{x,0\}$ for any $x\in \real$.
\end{prop}
\begin{proof}
The main observation is that to end the Phase II and begin the Phase III in the immediate next slot, one must make the estimated sparsity non-increasing. By step 11 of MCHTP in Table~\ref{algo:MCHTP}, this can occur at slot $t$, if $\Delta E_t<\epsilon$. However, under the conditions \eqref{eq:delta-condition-lower-bound} and \eqref{eq:epsilon-condition-phaseI}, $\Delta E_t\ge \epsilon$, whenever $\min_{i\in \{0,1\}}K_{t,i}<K$. Since $K_{t-1}\ge K$ throughout after the end of Phase I, it is therefore ensured that after the Phase II ends, $\{K_t\}$ remains unchanged whenever $\min_{i\in \{0,1\}}K_{t,i}<K$. Consequently, it remains to investigate bounds which ensure that $\Delta E_t<\epsilon$, whenever $\min_{i\in \{0,1\}}K_{t,i}\ge K$. The proof follows by deriving a bound ensuring this condition. For a detailed proof, refer to Appendix~\ref{sec:proof-prop-phase2-duration}.
\end{proof}
The phase III of MCHTP commences when phase II ends. With the satisfaction of the conditions~\eqref{eq:delta-condition-lower-bound} and \eqref{eq:epsilon-condition-phaseI} by virtue of Proposition~\ref{prop:phase-2-duration}, the absolute error difference is sufficiently small, which in turn ensures that the estimated sparsity sequence is non-increasing. We say that the phase III \textit{stops}, when the estimated sparsity sequence converges to $K$. Then, it is immediate that the phase III stops whenever the randomly chosen sparsity at a time step is equal to $K$. This allows us to easily estimate the duration $T_3$ of phase III as stated below:
\begin{lem}
The phase III of MCHTP has duration $T_3$, where,
\begin{align}
\label{eq:T_3-charaterization}
\prob{T_3=t} & = r(1-r)^{t-1}, t\ge 1,
\end{align}
where $r = \frac{1}{\bar{K}-1}$.	
\end{lem}
\begin{proof}
Note that the phase III ends as soon as, following Phase II, one encounters $K_{t,1}=K$. Specifically, $T_3$ can be precisely defined to be the following:
\begin{align}
    \label{eq:T3-def}
    \ & T_3 = \min\bigg\{t\ge 1: K_{\tau,1}\ne K, K_{\tau + 1, 1}\ne K,\cdots,\nonumber\\
    \ & K_{t-1,1}\ne K, K_{t,1}=K\bigg\}.
\end{align}
Since the sequence $K_{\tau,1},\cdots K_{t,1}$ constitute a DTMC, it follows that, 
\begin{align}
    \prob{T_3=t} & = \prob{K_{\tau,1}\ne K,\cdots, K_{t-1,1}\ne K, K_{t,1}=K}\nonumber\\
    \ & = \left(\frac{\bar{K}-2}{\bar{K}-1}\right)^{t-1}\left(\frac{1}{\bar{K}-1}\right)\nonumber\\
    \ & = (1-r)^{t-1}r,
\end{align}
where, $r = \frac{1}{\bar{K}-1}$.
\end{proof}
Consequently, the expected \textit{waiting time} $W$, to estimate the correct sparsity is obtained as below:
\begin{align}
\lefteqn{\expect{W}  = \expect{T_1}+\expect{T_2}+\expect{T_3}} & &\nonumber\\
\ & \le \frac{p}{1-q} + \left\lfloor\relu{\frac{\ln\left[\frac{\sqrt{(1-\delta)\epsilon}}{(1+\delta)\norm{\bvec{x}}}\right]}{\ln \rho}}\right\rfloor + \frac{1}{r}\nonumber\\
\ & = \left(1-\frac{1}{\bar{K}}\right)\frac{K-1}{\bar{K}-K+1} + \left\lfloor\relu{\frac{\ln\left[\frac{\sqrt{(1-\delta)\epsilon}}{(1+\delta)\norm{\bvec{x}}}\right]}{\ln \rho}}\right\rfloor\nonumber\\
\ & + \bar{K} - 1.
\end{align}
\section{Numerical Experiments}
\label{sec:sims}
\begin{figure}[t!]
\begin{subfigure}{0.5\textwidth}
\centering
    \includegraphics[scale=0.4]{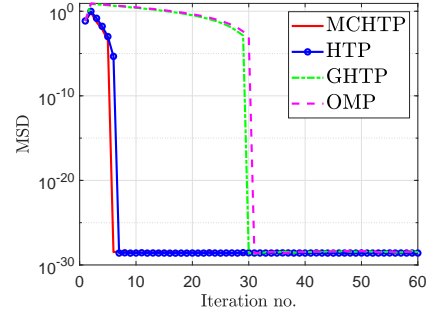}
    \subcaption{$K=30,\ K_{\max}=M/2$}
\end{subfigure}
\begin{subfigure}{0.5\textwidth}
\centering
    \includegraphics[scale=0.4]{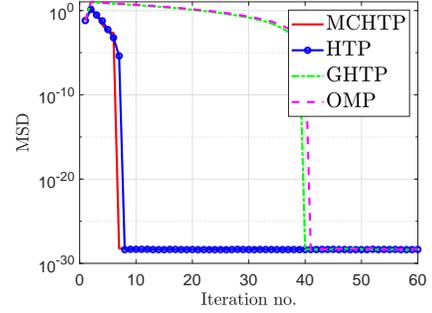}
    \subcaption{$K=40,\ K_{\max}=M/2$}
\end{subfigure}
    \caption{MSD vs iteration no. for MCHTP, HTP and GHTP.}
    \label{fig:msd-vs-time}
\end{figure}
\begin{figure}[t!]
    \centering
    \begin{subfigure}{0.5\textwidth}
    \centering
        \includegraphics[scale=0.4]{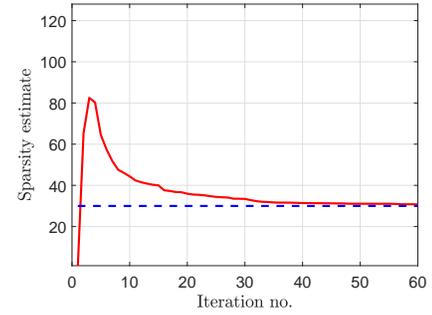}
        \subcaption{$K=30,\ K_{\max}=M/2$}
    \end{subfigure}
    \begin{subfigure}{0.5\textwidth}
    \centering
        \includegraphics[scale=0.4]{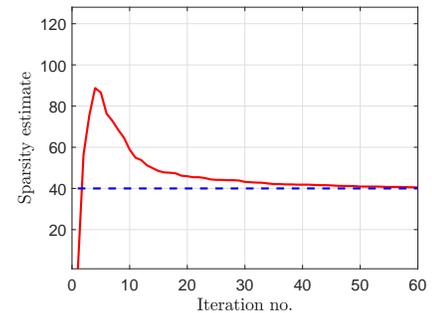}
        \subcaption{$K=40,\ K_{\max}=M/2$}
    \end{subfigure}
    \caption{Estimated sparsity sequence of MCHTP vs iteration no.}
    \label{fig:sparsity-vs-time}
\end{figure}
In this section we numerically evaluate both the sparse recovery and sparsity order estimation performance of MCHTP and compare it with other state-of-the-art results. For this purpose we have chosen the parameters $N=512, M=256, \mu = 0.3$. We have fixed $K_{\max} = M/2$, which is a reasonable bound since theoretically perfect recovery for a $K$ sparse vector is possible only if $K\le M/2$. Each experiment is repeated over $50$ independent instances and the results are averaged.
\section{Conclusion}
\label{sec:conclusion}
In this paper we have proposed a novel greedy sparse recovery algorithm referred to as MCHTP that can simultaneously recover an unknown sparse vector along with its unknown sparsity by using a novel decision criterion and HTP calculations at each iteration. Assuming noiseless measurements, we have provided theoretical analysis of the evolution of the sparsity sequence through phases and have provided theoretical bounds on the sensing matrix as well as the signal structure, which ensure perfect recovery of the vector as well as its unknown sparsity. Finally, the theoretical results are corroborated with numerical simulations that demonstrate the superior convergence of MCHTP as well as its unique sparsity estimation capability compared to the popular techniques.
\bibliography{tchtp_bib}
\appendices
\section{Proof of Proposition~\ref{prop:x^t-x^t-i-decay-inequalities}}
\label{sec:appendix-proof-prop-x^t-x^t-i-decay-inequalities}
We revisit the analysis of Theorem 3.5 of~. One can verify that the first step of the proof, i.e., the analysis up to Eq.~(3.11) in therein is applicable here without any modification and results in the following:
\begin{align}
\norm{\bvec{x}^t_i - \bvec{x}}^2 & \le
\label{eq:step1-upper-bound} \frac{1}{1-\delta_{K_{t,i}+K}^2}\norm{\big(\bvec{x}^t_i - \bvec{x}\big)_{\overline{\Lambda_{t,i}}}}^2. 
\end{align} 
On the other hand, applying the second step of the proof requires a little more care. Let us define $\Gamma_{t,i}$ as the support corresponding to the (magnitude-wise) top $K_{t,i}$ entries of $\bvec{x}$. We note that 
\begin{align}
\lefteqn{\norm{(\bvec{x}^{t-1}+\bvec{\Phi}^\top(\bvec{y}-\bvec{\Phi x}^{t-1}))_{\Gamma_{t,i}\setminus \Lambda_{t,i}}}} & &\nonumber\\
\label{eq:intermediate-step2}
\ & \le \norm{(\bvec{x}^{t-1}+\bvec{\Phi}^\top(\bvec{y}-\bvec{\Phi x}^{t-1}))_{\Lambda_{t,i}\setminus \Gamma_{t,i}}}.
\end{align}
We will now separately obtain lower and upper bounds of the left hand side (LHS) and the right hand side (RHS), respec., of the inequality~\eqref{eq:intermediate-step2}.

First note that the RHS of inequality~\eqref{eq:intermediate-step2} can be upper bounded as below:
\begin{align}
\lefteqn{\norm{(\bvec{x}^{t-1}+\bvec{\Phi}^\top(\bvec{y}-\bvec{\Phi x}^{t-1}))_{\Lambda_{t,i}\setminus \Gamma_{t,i}}}} & & \nonumber\\
\ & = \norm{(\bvec{x} + (\bvec{I}-\bvec{\Phi}^\top \bvec{\Phi})(\bvec{x}^{t-1}-\bvec{x}))_{\Lambda_{t,i}\setminus \Gamma_{t,i}}} \nonumber\\
\label{eq:step2-rhs-upper-bound-temp}
\ & \le \norm{\bvec{x}_{\Lambda_{t,i}\setminus \Gamma_{t,i}}} + \norm{((\bvec{I}-\bvec{\Phi}^\top \bvec{\Phi})(\bvec{x}^{t-1}-\bvec{x}))_{\Lambda_{t,i}\setminus \Gamma_{t,i}}}.
\end{align}
On the other hand, the LHS of inequality~\eqref{eq:intermediate-step2} can be lower bounded as below:
\begin{align}
\lefteqn{\norm{(\bvec{x}^{t-1}+\bvec{\Phi}^\top(\bvec{y}-\bvec{\Phi x}^{t-1}))_{\Gamma_{t,i}\setminus \Lambda_{t,i}}}} & & \nonumber\\
\ & = \norm{\big(\bvec{x}^t_i - \bvec{x}-(\bvec{I}-\bvec{\Phi}^\top \bvec{\Phi})(\bvec{x}^{t-1}-\bvec{x})}\nonumber\\
\ & \ge \norm{\big(\bvec{x}^t_i - \bvec{x}\big)_{\Gamma_{t,i}\setminus \Lambda_{t,i}}}\nonumber\\
\label{eq:step2-lhs-lower-bound-temp}
\ & - \norm{\big((\bvec{I}-\bvec{\Phi}^\top \bvec{\Phi})(\bvec{x}^{t-1}-\bvec{x})\big)_{\Gamma_{t,i}\setminus \Lambda_{t,i}}}.
\end{align}
Combining the bounds~\eqref{eq:step2-rhs-upper-bound-temp} and~\eqref{eq:step2-lhs-lower-bound-temp}, one obtains, 
\begin{align}
\lefteqn{\norm{\big(\bvec{x}^t_i - \bvec{x}\big)_{\Gamma_{t,i}\setminus \Lambda_{t,i}}}} & & \nonumber\\
\ & \le \sqrt{2}\norm{\big((\bvec{I}-\bvec{\Phi}^\top \bvec{\Phi})(\bvec{x}^{t-1}-\bvec{x})\big)_{\Gamma_{t,i}\Delta \Lambda_{t,i}}}\nonumber\\
\ & + \norm{\bvec{x}_{\Lambda_{t,i}\setminus \Gamma_{t,i}}}.
\end{align}
Therefore,
\begin{align}
\lefteqn{\norm{\big(\bvec{x}^t_i - \bvec{x}\big)_{\overline{\Lambda_{t,i}}}}} & & \nonumber\\
\ & \le \norm{\big(\bvec{x}^t_i - \bvec{x}\big)_{\Gamma_{t,i}\setminus \Lambda_{t,i}}} + \norm{\bvec{x}_{\overline{\Gamma_{t,i}\cup \Lambda_{t,i}}}}\nonumber\\
\ & \le \sqrt{2}\norm{\big((\bvec{I}-\bvec{\Phi}^\top \bvec{\Phi})(\bvec{x}^{t-1}-\bvec{x})\big)_{\Gamma_{t,i}\Delta \Lambda_{t,i}}}\nonumber\\
\ & + \norm{\bvec{x}_{\Lambda_{t,i}\setminus \Gamma_{t,i}}} + \norm{\bvec{x}_{\overline{\Gamma_{t,i}\cup \Lambda_{t,i}}}} \nonumber\\
\ & \le \sqrt{2}\norm{\big((\bvec{I}-\bvec{\Phi}^\top \bvec{\Phi})(\bvec{x}^{t-1}-\bvec{x})\big)_{\Gamma_{t,i}\Delta \Lambda_{t,i}}} + \sqrt{2}\norm{\bvec{x}_{\overline{\Gamma_{t,i}}}}\nonumber\\
\label{eq:step2-upper-bound}
\ & \le \sqrt{2}\delta_{K_{t,i}+K_{t-1}+K}\norm{\bvec{x}^{t-1}-\bvec{x}} + \sqrt{2}\norm{\bvec{x}_{\overline{\Gamma_{t,i}}}}.
\end{align}
Combining the bounds~\eqref{eq:step1-upper-bound} and~\eqref{eq:step2-upper-bound}, we obtain the following:
\begin{align}
\label{eq:decay-inequality}
\norm{\bvec{x}^t_i - \bvec{x}} & \le \rho_{t,i}\norm{\bvec{x}^{t-1}-\bvec{x}} + \gamma_{t,i}\norm{\bvec{x}_{\overline{\Gamma_{t,i}}}},
\end{align}
where $\rho_{t,i} = \frac{\sqrt{2}\delta_{K_{t,i}+K_{t-1}+K}}{\sqrt{1-\delta_{K_{t,i}+K}^2}}$ and $\gamma_{t,i} = \frac{\sqrt{2}}{\sqrt{1-\delta_{K_{t,i}+K}^2}}$. 

It is now clear that the following holds true:
\begin{align}
\label{eq:decay-inequality-xt}
\norm{\bvec{x}^t - \bvec{x}} & \le \rho_{t}\norm{\bvec{x}^{t-1}-\bvec{x}} + \gamma_{t}\norm{\bvec{x}_{\overline{\Gamma_{t}}}},
\end{align}
where $\rho_{t} = \frac{\sqrt{2}\delta_{K_{t}+K_{t-1}+K}}{\sqrt{1-\delta_{K_{t}+K}^2}}$ and $\gamma_{t} = \frac{\sqrt{2}}{\sqrt{1-\delta_{K_{t}+K}^2}}$,
and $\Gamma_{t}$ is the support corresponding to the (magnitude-wise) top $K_{t}$ entries of $\bvec{x}$.
\section{Proof of Proposition~\ref{prop:phase-I-MCHTP}}
\label{sec:appendix-proof-prop-phase-I-MCHTP}
%

Let us denote, at time $t\ge 0$, $i=\argmin_{j\in \{0,1\}}K_{t,j},\ l = \argmax_{j\in \{0,1\}}K_{t,j}$. The corresponding estimates produced are $\bvec{x}^t_i$ and $\bvec{x}^{t}_l$ with the respective supports $\Lambda_{t,i}$ and $\Lambda_{t,l}$ and errors $E_{t,i}=\norm{\dualproj{\Lambda_{t,i}}\bvec{y}}^2$ and $E_{t,l}=\norm{\dualproj{\Lambda_{t,l}}\bvec{y}}^2$. Step 4 of MCHTP guarantees that $\Lambda_{t,i}\subset \Lambda_{t,l}$. Let us denote $\Omega_{t}=\Lambda_{t,l}\setminus \Lambda_{t,i}$. Then one finds that,
\begin{align}
\norm{\dualproj{\Lambda_{t,l}}\bvec{y}}^2 & = \norm{\dualproj{\Lambda_{t,i}}\bvec{y}}^2 - \norm{\widetilde{\proj{\Omega_{t}}}\bvec{y}}^2,
\end{align}
where 
\begin{align}
\label{eq:tilde-projection}
    \widetilde{\proj{\Omega_{t}}} & =\dualproj{\Lambda_{t,i}}\bvec{\Phi}_{\Omega_{t}}(\dualproj{\Lambda_{t,i}}\bvec{\Phi}_{\Omega_{t}})^\dagger\nonumber\\
    \ & =\dualproj{\Lambda_{t,i}}\bvec{\Phi}_{\Omega_{t}}(\bvec{\Phi}_{\Omega_{t}}^\top\dualproj{\Lambda_{t,i}}\bvec{\Phi}_{\Omega_{t}})^{-1}\bvec{\Phi}_{\Omega_{t}}^\top\dualproj{\Lambda_{t,i}}
\end{align} Let us denote, $\Delta E_t = E_{t,i}-E_{t,l}$. Then, it is clear that $\Delta E_t = \norm{\widetilde{\proj{\Omega_{t}}}\bvec{y}}^2$. We will now obtain a lower bound on $\Delta E_{t}$.

First note that, using the lower bound on the eigenvalues of $\dualproj{\Lambda_{t,i}}\bvec{\Phi}_{\Omega_{t}}(\bvec{\Phi}_{\Omega_{t}}^\top\dualproj{\Lambda_{t,i}}\bvec{\Phi}_{\Omega_{t}})^{-1}\bvec{\Phi}_{\Omega_{t}}^\top$, one obtains, 
\begin{align}
\Delta E_t & = \norm{\dualproj{\Lambda_{t,i}}\bvec{\Phi}_{\Omega_{t}}(\bvec{\Phi}_{\Omega_{t}}^\top\dualproj{\Lambda_{t,i}}\bvec{\Phi}_{\Omega_{t}})^{-1}\bvec{\Phi}_{\Omega_{t}}^\top\dualproj{\Lambda_{t,i}}\bvec{y}}^2 \nonumber\\
\ & \ge \frac{\norm{\bvec{\Phi}_{\Omega_{t}}^\top\dualproj{\Lambda_{t,i}}\bvec{y}}^2}{(1+\delta_{K_{t,l}})}.
\end{align}
We will now derive a lower bound of $\norm{\bvec{\Phi}_{\Omega_{t}}^\top\dualproj{\Lambda_{t,i}}\bvec{y}}$. In order to do so, first observe that $\dualproj{\Lambda_{t,i}}\bvec{y} = \bvec{y} - \bvec{\Phi}\bvec{x}^t_i$. Using this, adding and subtracting $\bvec{x}^{t-1}$ and using the reverse triangle inequality, one obtains, 
\begin{align}
\norm{\bvec{\Phi}_{\Omega_{t}}^\top\dualproj{\Lambda_{t,i}}\bvec{y}} & \ge \norm{(\bvec{x}^{t-1}+\bvec{\Phi}^\top(\bvec{y}-\bvec{\Phi x}^{t-1}))_{\Omega_t}}\nonumber\\
\ & - \norm{((\bvec{I}-\bvec{\Phi}^\top\bvec{\Phi})(\bvec{x}^t_i-\bvec{x}^{t-1}))_{\Omega_t}},
\end{align}
where in the last step we have used the fact that $[\bvec{x}^t_i]_j = 0,\ \forall j\in \Omega_t$. Now observe that the step 4 of MCHTP (Algorithm~\ref{algo:MCHTP}) ensures
that the sets $\Lambda_{t,i}$ and $\Lambda_{t,l}$ contain the top (magnitude wise) $K_{t,i}$ and $K_{t,l}$ entries, respectively, of the vector $\bvec{x}^{t-1}+\bvec{\Phi}^\top(\bvec{y-\bvec{\Phi x}^{t-1}})$. Now depending on whether or not $K_{t,l}-K_{t,i}> K$, there are two cases to consider. If $K_{t,l}-K_{t,i}\le K$, then, for all $j=1,2,\cdots, K_{t,l}-K_{t,i}$, the $j^\mathrm{th}$ maximum of the vector $\bvec{x}^{t-1}+\bvec{\Phi}^\top(\bvec{y-\bvec{\Phi x}^{t-1}})$ from the set $\Omega_t$ is greater than or equal to the $j^\mathrm{th}$ maximum of $\abs{(\bvec{x}^{t-1}+\bvec{\Phi}^\top(\bvec{y}-\bvec{\Phi x}^{t-1}))}$ from the set $\Lambda\setminus \Lambda_{t,i}$. On the other hand, if $K_{t,l}-K_{t,i}> K$, then all the entries of the corresponding vector from the set $\Omega_t$ are larger than the entries from the set $\Lambda\setminus \Lambda_{t,i}$. Therefore, in any case, we have the following: 
\begin{align}
\lefteqn{\norm{(\bvec{x}^{t-1}+\bvec{\Phi}^\top(\bvec{y}-\bvec{\Phi x}^{t-1}))_{\Omega_t}}} & & \nonumber\\
\ & \ge \norm{(\bvec{x}^{t-1}+\bvec{\Phi}^\top(\bvec{y}-\bvec{\Phi x}^{t-1}))_{{\Gamma}_{t,l,i}}}\nonumber\\
\ & \ge \norm{\bvec{\Phi}_{{\Gamma}_{t,l,i}}^\top(\bvec{y}-\bvec{\Phi x}_i^t))}\nonumber\\
\ & -\norm{((\bvec{I}-\bvec{\Phi}^\top\bvec{\Phi})(\bvec{x}_i^t-\bvec{x}^{t-1}))_{{\Gamma}_{t,l,i}}},
\end{align}
where $\Gamma_{t,l,i}$ is the subset of $\Lambda\setminus \Lambda_{t,i}$ which contains the $K_{t,l,i}:=\min\{K_{t,l}-K_{t,i},K\}$ (magnitudewise) largest entries of $\bvec{x}^{t-1}+\bvec{\Phi}^\top(\bvec{y}-\bvec{\Phi x}^{t-1})$. We have furthermore used the fact that $[\bvec{x}^t_i]_j = 0,\ \forall j\in \Lambda\setminus\Lambda_{t,i}$.Therefore, we obtain, 
\begin{align}
\lefteqn{\norm{\bvec{\Phi}_{\Omega_t}^\top\dualproj{\Lambda_{t,i}}\bvec{y}}} & &\nonumber\\
\ & \ge \norm{\bvec{\Phi}_{\Gamma_{t,l,i}}^\top(\bvec{y}-\bvec{\Phi x}_i^t))}\nonumber\\
\ & -\norm{((\bvec{I}-\bvec{\Phi}^\top\bvec{\Phi})(\bvec{x}_i^t-\bvec{x}^{t-1}))_{\Gamma_{t,l,i}}}\nonumber\\
\ & - \norm{((\bvec{I}-\bvec{\Phi}^\top\bvec{\Phi})(\bvec{x}_i^t-\bvec{x}^{t-1}))_{\Omega_t}}\nonumber\\
\ & \ge \norm{\bvec{\Phi}_{\Gamma_{t,l,i}}^\top(\bvec{y}-\bvec{\Phi x}_i^t))}\nonumber\\
\ & - \sqrt{2}\norm{\bigg((\bvec{I}-\bvec{\Phi}^\top\bvec{\Phi})(\bvec{x}_i^t-\bvec{x}^{t-1})\bigg)_{\Omega_t\cup \Gamma_{t,l,i}}}\nonumber\\
\ & \ge \norm{\bvec{\Phi}_{\Gamma_{t,l,i}}^\top(\bvec{y}-\bvec{\Phi x}_i^t))}\nonumber\\
\ & - \sqrt{2}\delta_{K_{t,l}-K_{t,i}+K_{t-1}+K}\norm{\bvec{x}_i^t-\bvec{x}^{t-1}}. 
\end{align}
Now note that 
\begin{align}
\lefteqn{\norm{\bvec{\Phi}_{\Gamma_{t,l,i}}^\top(\bvec{y}-\bvec{\Phi x}_i^t))}} & & \nonumber\\
\ & \ge \frac{\sqrt{\abs{\Gamma_{t,l,i}}}\norm{((\bvec{\Phi}^\top(\bvec{y}-\bvec{\Phi}\bvec{x}_i^t)))_{\Lambda\setminus \Lambda_{t,i}}}}{\sqrt{\abs{\Lambda\setminus \Lambda_{t,i}}}}\nonumber\\
\ & \stackrel{(a)}{\ge} \frac{\norm{((\bvec{\Phi}^\top(\bvec{y}-\bvec{\Phi}\bvec{x}_i^t)))_{\Lambda\setminus \Lambda_{t,i}}}}{\sqrt{K}}\nonumber\\
\ & = \frac{\norm{((\bvec{\Phi}^\top\bvec{\Phi}(\bvec{x}_i^t-\bvec{x})))_{\Lambda\setminus \Lambda_{t,i}}}}{\sqrt{K}}\nonumber\\
\ & \stackrel{(b)}{=}\frac{\norm{((\bvec{\Phi}^\top\bvec{\Phi}(\bvec{x}_i^t-\bvec{x})))_{\Lambda\cup \Lambda_{t,i}}}}{\sqrt{K}}\nonumber\\
\ & \ge \frac{(1-\delta_{K_{t,i}+K})\norm{\bvec{x}_i^t-\bvec{x}}}{\sqrt{K}},
\end{align}
where the step $(a)$ follows from the fact that $\abs{\Lambda\setminus \Lambda_{t,i}}\le \abs{\Lambda}=K$ and $\abs{\Gamma_{t,l,i}}=K_{t,l,i}=\min\{K_{t,l}-K_{t,i},K\}\ge 1$. Furthermore, step $(b)$ follows from the fact that $\bvec{\Phi}^\top_{\Lambda_{t,i}}(\bvec{y}-\bvec{\Phi x}^t_i)=\bvec{0}_{\Lambda_{t,i}}$. Moreover,
\begin{align}
\norm{\bvec{x}_i^t-\bvec{x}} & \ge \min_{\substack{\bvec{z}\in \real^N:\\ \opnorm{\bvec{z}}{0}\le K_{t,i}}} \norm{\bvec{z}-\bvec{x}} = \norm{\bvec{x}_{\overline{\Gamma}_{t,i}}}.
\end{align}
On the other hand, using Eq.~\eqref{eq:decay-inequality}, we obtain,
\begin{align}
\norm{\bvec{x}^t_i-\bvec{x}^{t-1}} & \le \norm{\bvec{x}^t_i-\bvec{x}} + \norm{\bvec{x}^{t-1}-\bvec{x}}.
\end{align}
Therefore, we obtain,
\begin{align}
    \lefteqn{\norm{\bvec{\Phi}_{\Omega_t}^\top\dualproj{\Lambda_{t,i}}\bvec{y} }} & &\nonumber\\
\ & \ge \left(\frac{1-\delta_{K_{t,i}+K}}{\sqrt{K}} - \sqrt{2}\delta_{K_{t,l}-K_{t,i}+K_{t-1}+K}\right) \norm{\bvec{x}_{\overline{\Gamma}_{t,i}}}\nonumber\\
\ & - \sqrt{2}\delta_{K_{t,l}-K_{t,i}+K_{t-1}+K}\norm{\bvec{x}^{t-1}-\bvec{x}}.
\end{align}
Note that $K_{t,i}\ge 0,\ K_{t,l},K_{t-1}\le \bar{K}$. Therefore, for simplicity, we denote $\delta: = \delta_{2\bar{K}+K}$ 
and further lower bound the above as
\begin{align}
    \lefteqn{\norm{\bvec{\Phi}_{\Omega_t}^\top\dualproj{\Lambda_{t,i}}\bvec{y}} \ge \left(\frac{1-\delta}{\sqrt{K}} - \sqrt{2}\delta\right) \norm{\bvec{x}_{\overline{\Gamma}_{t,i}}}} & &\nonumber\\
    \label{eq:temp-bound1}
\ & - \sqrt{2}\delta\norm{\bvec{x}^{t-1}-\bvec{x}}.
\end{align}

 We now use the bound~\eqref{eq:decay-inequality-xt} recursively, and the fact that $\bvec{x}_0=\bvec{0}$, to obtain the following upper bound of $\norm{\bvec{x}^{t-1}-\bvec{x}}$:
\begin{align}
\label{eq:xt-1-bound}
    \norm{\bvec{x}^{t-1}-\bvec{x}} & \le \rho^{t-1} \norm{\bvec{x}} + \gamma \sum_{j=1}^{t-1}\rho^{t-1-j}\norm{\bvec{x}_{\overline{\Gamma}_{j}}},
\end{align}
where we have used the bound~\eqref{eq:x^t-decay-inequality} of Proposition~\ref{prop:x^t-x^t-i-decay-inequalities} and have denoted $\rho = \frac{\sqrt{2}\delta}{\sqrt{1-\delta^2}},\ \gamma = \frac{\sqrt{2}}{\sqrt{1-\delta^2}}$
Therefore, taking the bounds~\eqref{eq:temp-bound1} and~\eqref{eq:xt-1-bound}, we obtain
\begin{align}
    \lefteqn{\norm{\bvec{\Phi}_{\Omega_t}^\top\dualproj{\Lambda_{t,i}}\bvec{y}} \ge \left(\frac{1-\delta}{\sqrt{K}} - \sqrt{2}\delta\right) \norm{\bvec{x}_{\overline{\Gamma}_{t,i}}}} & & \nonumber\\
    \label{eq:general-lower-bound}
    \ & - \sqrt{2}\delta\bigg(\rho^{t-1} \norm{\bvec{x}} + \gamma \sum_{j=1}^{t-1}\rho^{t-1-j}\norm{\bvec{x}_{\overline{\Gamma}_{j}}}\bigg).
\end{align}
Consider the first step of MCHTP. In this step, $t=1$ and $\bvec{x}^0=\bvec{0},\Gamma_0=\emptyset$ since $K_0=0$, so that $\bvec{x}_{\overline{\Gamma}_{0}}=\norm{\bvec{x}}$. Furthermore, $K_{1,i}=K_0=0$. Therefore, the bound~\eqref{eq:temp-bound1} reduces to \begin{align}
\label{eq:lower_bound_t=1}
    \norm{\bvec{\Phi}_{\Omega_1}^\top\dualproj{\Lambda_{1,i}}\bvec{y}} & \ge \left(\frac{1-\delta}{\sqrt{K}} - 2\sqrt{2}\delta\right) \norm{\bvec{x}}.
\end{align} 
Therefore, at $t=1$, $K_1=K_{1,l}$ if $\Delta E_1\ge \epsilon$, which always holds if 
\begin{align}
\label{eq:condition-step=1}
    \left(\frac{1-\delta}{\sqrt{K}} - 2\sqrt{2}\delta\right) \norm{\bvec{x}}  & \ge \sqrt{\epsilon(1+\delta)}.
\end{align} 
Obviously, we require, $\frac{1-\delta}{\sqrt{K}} - 2\sqrt{2}\delta>0$, which is satisfied by the following:
\begin{align}
    \label{eq:lower-bound-t=1-cond1}
    \delta < \frac{1}{2\sqrt{2K}+1}.
\end{align}
Therefore, the conditions~\eqref{eq:condition-step=1} and \eqref{eq:lower-bound-t=1-cond1} ensure that $\Delta E_1>0$ so that $K_{1}=K_{1,l}$.

Now consider $t>1$. Using the fact that $\Gamma_{t,i}\subset \Gamma_{t-1}$ since $ K_{t,i}\le K_{t-1}$, we obtain from the bound~\eqref{eq:general-lower-bound} the following:
\begin{align}
    \lefteqn{\norm{\bvec{\Phi}_{\Omega_t}^\top\dualproj{\Lambda_{t,i}}\bvec{y}} \ge \left(\frac{1-\delta}{\sqrt{K}} - \sqrt{2}\delta(1+\gamma)\right) \norm{\bvec{x}_{\overline{\Gamma}_{t,i}}}} & & \nonumber\\
        \label{eq:general-lower-bound-t>1-prelim}
    \ & - \sqrt{2}\delta\bigg(\rho^{t-1} \norm{\bvec{x}} + \rho\gamma \sum_{j=1}^{t-2}\rho^{t-2-j}\norm{\bvec{x}_{\overline{\Gamma}_{j}}}\bigg).
\end{align}
For any $t>1$, we can use the bounds $\norm{\bvec{x}_{\overline{\Gamma}_t}}\le\sqrt{1-\frac{K_{t}}{K}}\norm{\bvec{x}}\le \sqrt{1-\frac{1}{K}}\norm{\bvec{x}}$, since $K_t\ge 1$ and $\norm{\bvec{x}_{\overline{\Gamma}_{t,i}}}\ge \sqrt{K-K_{t,i}}x_{\min}\ge x_{\min}$ since $K_{t,i}\le K-1$. Furthermore, assuming $\delta<\frac{1}{\sqrt{2K}+1}<\frac{1}{\sqrt{3}}$, we have $\rho<1$ (since we $K\ge 1$) (we will later obtain conditions which indeed implies this bound on $\delta$). With the above observations, we simplify the lower bound of~\eqref{eq:general-lower-bound-t>1-prelim} to the following,
\begin{align}
    \label{eq:temp-lower_bound_t>1}
    \lefteqn{\norm{\bvec{\Phi}_{\Omega_t}^\top\dualproj{\Lambda_{t,i}}\bvec{y}}} & & \nonumber\\
    \ & \ge x_{\min}\left(\frac{1-\delta}{\sqrt{K}} - \sqrt{2}\delta(1+\gamma)\right) \nonumber\\
    \ & - \sqrt{2}\delta\norm{\bvec{x}}\bigg(\rho^{t-1} + \frac{\rho\gamma\sqrt{1-\frac{1}{K}}(1-\rho^{t-2})}{1-\rho}\nonumber\\
    \ & + \frac{\rho\norm{\bvec{e}}}{(1-\rho)\norm{\bvec{x}} }\bigg).
\end{align}
Now, note that using the bound $\delta<\frac{1}{\sqrt{2K+1}}$, we obtain, for $K\ge 2$,
\begin{align}
    \frac{\gamma\sqrt{1-\frac{1}{K}}}{1-\rho} & = \frac{\sqrt{2}\sqrt{1-\frac{1}{K}}}{\sqrt{1-\delta^2}-\sqrt{2}\delta}\nonumber\\
    \ & \le \frac{\sqrt{2K+1}\sqrt{K-1}}{\sqrt{K}(\sqrt{K}-1)}< 4,
\end{align}
and 
\begin{align}
    \frac{1}{1-\rho} = \frac{\sqrt{1-\delta^2}}{\sqrt{1-\delta^2}-\sqrt{2}\delta} & \le \frac{\sqrt{K}}{\sqrt{K}-1}\le 2 + \sqrt{2}<4,
\end{align}
and,
\begin{align}
    \gamma & = \frac{\sqrt{2}}{\sqrt{1-\delta^2}} \le \frac{\sqrt{2K+1}}{\sqrt{K}}\le \sqrt{2.5}<2,\nonumber\\
    \rho & = \frac{\sqrt{2}\delta}{\sqrt{1-\delta^2}} \le \frac{\sqrt{2}}{\sqrt{2K}} = \frac{1}{\sqrt{K}}.
\end{align}
Therefore from~\eqref{eq:temp-lower_bound_t>1}, we obtain, for $t>1$, 
\begin{align}
    \lefteqn{\norm{\bvec{\Phi}_{\Omega_t}^\top\dualproj{\Lambda_{t,i}}\bvec{y}} \ge x_{\min}\left(\frac{1-\delta}{\sqrt{K-1}} - 3\sqrt{2}\delta\right)} & & \nonumber\\
    \ & - \sqrt{2}\delta\frac{\norm{\bvec{x}}}{\sqrt{K}}\bigg(\frac{1}{K^{t/2-1}} + 4 + \frac{4\norm{\bvec{e}}}{\norm{\bvec{x}}}\bigg),
\end{align}
which implies that,
\begin{align}
     \lefteqn{\norm{\bvec{\Phi}_{\Omega_t}^\top\dualproj{\Lambda_{t,i}}\bvec{y}} > x_{\min}\bigg(\frac{1-\delta}{\sqrt{K}} - 3\sqrt{2}\delta\bigg)} & &\nonumber\\
     \ &  - x_{\max}\sqrt{2}\delta\bigg(4+\frac{1}{K^{t/2-1}}\bigg). 
\end{align}
We can further lower bound the right hand side of the above inequality by using $K_{t-1}\le K - 1$ and $K^{t/2-1}\ge 1$ for $t\ge 2$, to obtain the following lower bound whenever $K_{t,i}<K$:
\begin{align}
    \label{eq:lower_bound_t>1}
     \norm{\bvec{\Phi}_{\Omega_t}^\top\dualproj{\Lambda_{t,i}}\bvec{y}} > x_{\min}\bigg(\frac{1-\delta}{\sqrt{K}} - 3\sqrt{2}\delta\bigg) &  - x_{\max}5\sqrt{2}\delta. 
\end{align}
Therefore, for $t>1$, if $K_{t,i}<K$, then $K_t=K_{t,l}$ if $\Delta E_t\ge \epsilon$, which always holds if 
\begin{align}
    \label{eq:condition-step-t>1}
    \ & x_{\min}\bigg(\frac{1-\delta}{\sqrt{K}} - 3\sqrt{2}\delta\bigg)  - 5\sqrt{2}\delta x_{\max}\ge \sqrt{\epsilon(1+\delta)}.
\end{align}
To make the LHS of~\eqref{eq:lower_bound_t>1} positive, we obviously require $x_{\min}\bigg(\frac{1-\delta}{\sqrt{K}} - 3\sqrt{2}\delta\bigg) - x_{\max}5\sqrt{2}\delta>0$, which is ensured by the following:
\begin{align}
    \label{eq:delta-bound-t>1-cond1}
    \delta & < \frac{1}{1+(3+5R)\sqrt{2K}},
\end{align}
where $R = \frac{x_{\max}}{x_{\min}}$. 
Therefore, the conditions~\eqref{eq:condition-step-t>1} and \eqref{eq:delta-bound-t>1-cond1} ensure that, for any $t>1$, if $K_{t,i}<K$, then $\Delta E_t>0$ and $K_t= K_{t,l}$.

Furthermore, observe that the condition~\eqref{eq:delta-bound-t>1-cond1} implies the conditions~\eqref{eq:lower-bound-t=1-cond1}.
Therefore, under the condition~\eqref{eq:delta-condition-lower-bound}, we have the following lower bound for $\Delta E_t$
\begin{align}
\Delta E_t & \ge  \frac{(1-\delta)}{(1+\delta)^2}\left(a_K(\delta)x_{\min} - b(\delta)x_{\max}\right)^2,
\end{align}
where $a_K(\delta),b(\delta)$ are defined in Eq.~\eqref{eq:a_K-b_K-def}. Therefore, the condition~\eqref{eq:epsilon-condition-phaseI} ensures that $\Delta E_t>\epsilon$, which in turn guarantees that $K_t = K_{t,l}$, i.e., the larger sparsity value is chosen for time step $t$.
\section{Proof of Proposition~\ref{prop:phase-2-duration}}
\label{sec:proof-prop-phase2-duration}
The Phase II continues until the sequence of estimated sparsity becomes a non-increasing sequence. Now, by the step 11 of MCHTP, if the current sparsity estimate is $K_{t-1}$ and $K_{t,1}$ is the sparsity sampled at time $t$, $K_t = \min\{K_{t-1}, K_{t,1}\}$ if $\Delta E_t<\epsilon$. However, if $K_{t,1}<K$, thanks to Proposition~\ref{prop:phase-I-MCHTP}, $\Delta E_t\ge \epsilon$ and $K_{t}=K_{t-1}$ since $K_{t-1}\ge K$ after Phase I. On the other hand, if $K_{t,1}\ge K$, then $K_t=\min\{K_{t-1}, K_{t,1}\}$ if one has $\Delta E_t< \epsilon$. Therefore, it is enough to find conditions which guarantee that $\Delta E_t<\epsilon$, whenever $K_{t,1}\ge K$, to ensure that the sequence $K_t$ after the end of Phase II, is a non-increasing sequence.

Now note that,
\begin{align}
\lefteqn{\sqrt{\Delta E_t} = \norm{\widetilde{\proj{\Omega_{t}}}\bvec{y}} = \norm{\widetilde{\proj{\Omega_{t}}}\dualproj{\Lambda_{t,i}}\bvec{y}}\le \norm{\dualproj{\Lambda_{t,i}}\bvec{y}}} & &\nonumber\\
\ & =\norm{\bvec{\Phi}(\bvec{x}} \le \sqrt{1+\delta}\norm{\bvec{x}^{t}_i - \bvec{x}}.
\end{align}
Note that Phase II of MCHTP ends at time $T_1+T_2-1$ and Phase III begins at time $T_1+T_2$ when $\Delta E_t\le \epsilon$ for all $t\ge T_1+T_2$, whenever $K_{t,1}\ge K$. This is ensured if the following is satisfied at such a time instant $t$: 
\begin{align}
\label{eq:phase3-beings}
\norm{\bvec{x}^{t}_i - \bvec{x}} & \le \frac{\sqrt{\epsilon}}{\sqrt{1+\delta}}.
\end{align}
Since $K_{t,1}\ge K$, and $K_{t-1}\ge K$ by Porposition~\ref{prop:phase-I-MCHTP}, we have $K_{t,i}\ge K$. Therefore, from Proposition~\ref{prop:x^t-x^t-i-decay-inequalities} we obtain
\begin{align}
\norm{\bvec{x}^{t}_i - \bvec{x}} & \le \rho \norm{\bvec{x}^{t-1}-\bvec{x}}\nonumber\\
\ & \le \rho^{t-T_1+1}\norm{\bvec{x}^{T_1-1}-\bvec{x}}\nonumber\\
\ & \le \rho^{t-T_1+1}\sqrt{\frac{1+\delta}{1-\delta}}\norm{\bvec{x}}.
\end{align}
Consequently, it follows that, if $T_2$ is the duration of Phase II of MCHTP, then $T_2\le \tau$, where,
\begin{align}
\tau & =\min\left\{t\ge 0:\rho^{t+1}\sqrt{\frac{1+\delta}{1-\delta}}\norm{\bvec{x}}\right.\nonumber\\
\ & \le \left.\frac{{\sqrt{\epsilon}}}{\sqrt{1+\delta}}\right\} = \min\left\{t\ge 0:\rho^{t+1}\le G\right\},
\end{align}
where 
\begin{align}
    G & = \frac{\sqrt{\epsilon(1-\delta)}}{(1+\delta)\norm{\bvec{x}}}
\end{align}
\end{document}